\documentstyle[preprint,tighten,aps]{revtex}

\begin{document}
\draft
\title{
Low-temperature effective potential of the Ising model.
}
\author{Andrea Pelissetto and Ettore Vicari}
\address{Dipartimento di Fisica dell'Universit\`a 
and I.N.F.N., I-56126 Pisa, Italy}

\date{\today}

\maketitle

\begin{abstract}
We study the low-temperature effective potential of the Ising model.
We evaluate the three-point and four-point zero-momentum renormalized
coupling constants that parametrize the expansion of
the effective potential near the coexistence curve.
These results 
are obtained by a constrained  analysis of the $\epsilon$-expansion
that uses accurate estimates for the two-dimensional Ising model. 
\medskip


\end{abstract}

\newpage
\newcommand{\N}{\hbox{{\rm I}\kern-.2em\hbox{\rm N}}}

\section{Introduction}
\label{introduction}

In statistical physics the effective potential corresponds to the 
free-energy density ${\cal F}$ as a function of 
the order parameter, which, for spin models, is the magnetization $M$.
The global minimum of the effective potential
determines the value of the order parameter,
that characterizes the phase of the model.
In the high-temperature (HT) or symmetric phase the minimum is 
unique with $M=0$. According to the Ginzburg-Landau theory, 
as the temperature decreases below the critical value,
the effective potential takes a double-well shape. The order parameter
does not vanish anymore and the system is in the low-temperature (LT)
or broken phase. 
Actually in the broken phase the double-well
shape is not correct because the effective potential must always
be convex.
In this phase it should present a flat region around the origin.
For a discussion see e.g. Refs.~\cite{OR,M-S} and references therein.

Recently many works have been devoted to the study of the 
effective potential --- or, equivalently, of the equation of state ---
in the limit of small magnetization for $O(N)$ models in the 
HT phase. Indeed, in this case, the effective potential admits 
a regular expansion around $M=0$ with coefficients that are related to
the zero-momentum $n$-point renormalized coupling constants.
Rather accurate estimates of these quantities have been obtained 
exploiting various field-theoretic
approaches (see e.g. \cite{effpot,G-Z,G-Z-2,Sokolov,T-W,Morris}) 
and lattice techniques (see e.g. \cite{B-C-g,Z-L-F,Tsypin,B-K,K-L}). 

In the LT phase the effective potential shows a complex behaviour
due to the Goldstone bosons and does not admit a regular expansion.
The only exception is the Ising model where the symmetry is 
discrete and therefore the Goldstone bosons are absent. 
In this case, if $M_0 = \lim_{H\to 0+} M(H)$, 
for $M > M_0$ (i.e. for $H>0$), the effective potential admits 
a regular expansion in powers of $(M - M_0)$.
The coefficients 
of this expansion are related to the LT zero-momentum $n$-point 
renormalized coupling constants $g_n $.
These quantities are harder to estimate than the corresponding
HT couplings and indeed they are known with much less precision. 
At present there are some estimates for $g_3$ and $g_4$. However
the uncertainty is often very large or difficult to 
estimate (especially for $g_4$)~\cite{Z-L-F,B-T-W,K-P,Tsypinbr}.
Surprisingly one finds the same situation for the two-dimensional 
Ising model (see Refs.~\cite{Z-L-F,K-P}).

In this paper we study the LT effective potential for the Ising model 
in two and three dimensions. We focus on the expansion in
powers of $(M - M_0)$ that 
we parametrize in terms of $w$ defined by
\begin{equation}
w^2\equiv \lim_{T\to T_c - }\,  \lim_{H\to 0}{\chi\over M^2\xi^d}
\label{defw}
\end{equation}
(where $\chi$ is the magnetic susceptibility and
$\xi$ the second-moment correlation length),
and of the ratios $v_j\equiv g_j /w^{j-2}$.
In two dimensions  $M$ is known exactly (see e.g.
\cite{Baxter}) while $\chi$ and
$\xi$ can be derived 
using the fact that the two-point function satisfies a 
Painlev\'e differential equation in the critical limit~\cite{Wuising}.
Therefore one can easily obtain high-precision estimates of $w$.
The ratios $v_j$ are estimated analyzing the LT expansion 
(to $O(u^{23})$ where $u=e^{-4/T}$) 
of the free-energy on the square lattice in the presence of 
an external constant magnetic field~\cite{B-E}.
In three dimensions good estimates of $w^2$ have been obtained from 
the analysis of its LT expansion~\cite{Z-L-F,ONgr2} and from Monte Carlo
simulations~\cite{C-H}. Field-theoretic calculations
of $w$~\cite{Munster,Munster2} 
are less precise, but perfectly consistent.
In order to estimate the constants $v_j$ we consider 
their $\epsilon$-expansion, that can be derived from the 
$\epsilon$-expansion of the equation of state~\cite{B-W-W-e2,W-Z,N-A}.
A constrained analysis of these $\epsilon$-series using the 
corresponding
two-dimensional results allows us to obtain estimates of 
$v_3$ and $v_4$ with satisfactory accuracy.
It is straightforward to obtain corresponding estimates for
$g_j= w^{j-2} v_j$.

The paper is organized as follows.
In Sec.~\ref{sec2} we introduce our notation and give 
some general formulae for 
the LT expansion of the effective potential near  the coexistence curve.
In Sec.~\ref{sec3} we compute $w$ and $v_j$ for various values of 
$j$ for the two-dimensional Ising model. In Sec.~\ref{sec4} we present
our analysis of the $\epsilon$-expansion of $v_{j}$,
and we compare our results with other approaches.
In Sec. \ref{sec5} we discuss the effective potential for the 
$O(N)$ model in the large-$N$ limit. We find logarithmic 
terms that have not been previously predicted. We discuss some possible 
interpretations.
In App.~\ref{appa} we report the LT expansion of the 
susceptibilities that are used in the computation of $v_j$ with $j \le 6$
for the two-dimensional Ising model.

\section{Expansion of the effective potential 
         near the coexistence curve}
\label{sec2}

The effective potential is related to the free energy of the 
model. Indeed, if $M\equiv\langle \phi\rangle$ is the magnetization and 
$H$ the magnetic field, one defines 
\begin{equation}
{\cal F} (M) = M H - {1\over V} \log Z(H),
\end{equation}
where $Z(H)$ is the partition function and the dependence on the
temperature $T$ is always understood in the notation. We will
be interested in the behaviour of ${\cal F} (M)$ near the coexistence
curve $t\equiv T-T_c<0$, $H=0$. For the Ising model, 
if $M_0=\lim_{H\to 0+} M(H)$, for $M>M_0$, we can expand 
${\cal F} (M)$ in powers of $M - M_0$. Explicitly we can write
\begin{equation}
{\cal F} (M) = {\cal F} (M_0) + \sum_{j=2}^\infty {1\over j!} a_j (M-M_0)^j.
\end{equation}
It is useful to express the coefficients $a_j$ in terms 
of renormalization-group invariant quantities. We therefore 
define a renormalized magnetization
\begin{equation}
\varphi^2 = {\xi(t,H=0)^2 M(t,H)^2\over \chi(t,H=0)}
\end{equation}
where $\xi$ is the second-moment correlation length
\begin{equation}
\xi^2 = {1\over 2d} {\int dx \ x^2 G(x) \over \int dx\ G(x)},
\end{equation}
and $G(x)$ is the connected two-point function. In terms of $\varphi$
we can rewrite
\begin{equation}
{\cal F}(\varphi) = 
 {\cal F}(\varphi_0)+{1\over 2} m^2(\varphi-\varphi_0)^2 + 
\sum_{j=3} m^{d-j(d-2)/2}
{1\over j!} g _{j} (\varphi-\varphi_0)^j.
\label{freeeng}
\end{equation}
Here $m=1/\xi$ and $g_j$ are functions of $t$ only. In field 
theory $\varphi$ is nothing but the zero-momentum renormalized field.
For $t\to 0-$ 
the quantities $g_j$ approach universal constants (that we indicate with the
same symbol) that represent the zero-momentum
$n$-point renormalized coupling constants. An even simpler parametrization 
can be obtained if we introduce
\begin{equation}
z \equiv {\varphi\over \varphi_0} = {M\over M_0}.
\label{defzeta}
\end{equation} 
Then we have 
\begin{equation}
{\cal F}(\varphi)-{\cal F}(\varphi_0) = {m^d \over w^2} B(z).
\label{dAZ}
\end{equation}
The function $B(z)$ has the following expansion
\begin{equation}
B(z) =   {1\over 2} (z-1)^2 + 
\sum_{j=3} {1\over j!} v_{j} (z-1)^j,
\label{BZ}
\end{equation}
where
\begin{equation}
v_{j} = {g _{j}\over w^{j-2}} .
\label{uj}
\end{equation}
The advantage of the expansion (\ref{BZ}) is that 
its coefficients $v_j$ are expressed only in terms of 
zero-momentum quantities, while in Eq. (\ref{freeeng}) 
the correlation length is also present.

We mention that a more natural expansion of $B(z)$ that takes 
into account its parity properties would be
\begin{equation}
B(z) = {1\over 4}  \left[ {1\over 2} (z^2-1)^2 + 
\sum_{j=3} {1\over j!} \overline{v}_{j} (z^2-1)^j\right],
\label{BZ2}
\end{equation}
where the coefficients $\overline{v}_j$ can be easily related to 
those of the expansion (\ref{BZ}), i.e. to the constants $v_j$.
For example $\overline{v}_3 = (3-v_3)/2$, etc....

Since $z\propto |t|^{-\beta} M$, 
the equation of state can be written in the form
\begin{equation}
H= \left({\partial {\cal F}\over \partial M}\right)_t
\propto |t|^{\beta\delta} {\partial B(z)\over \partial z}
\label{eqa}
\end{equation}
where we have used the hyperscaling relation $\beta(1+\delta)-d\nu=0$.
Eq. (\ref{eqa}) can be exploited to derive
$B(z)$ from the equation of state that is written in the form 
\begin{equation}
H=M^\delta f(x)
\label{eqstate}
\end{equation}
with $x=t M^{-{1/\beta}}$. The scaling function $f(x)$ is usually normalized
so that $f(-1) = 0$, $f(0) = 1$.
The function $B(z)$ can be obtained from
\begin{equation}
{\partial B(z)\over \partial z} = 
h_0 z^\delta f\left(x_0 z^{-{1/\beta}}\right),
\label{daz}
\end{equation}
where the normalization constants $h_0$ and $x_0$ are fixed by
the requirement that
\begin{equation}
B(z) = {1\over 2}(z-1)^2 + O\left[(z-1)^3\right].
\label{norm}
\end{equation}
Since $\beta>0$ and the function $f(x)$ is regular at $x=0$
and nonzero, Eq.~(\ref{daz}) implies that $B(z)\sim z^{\delta+1}$ for
$z\to \infty$.

The coefficients $v_j$ of the expansion of $B(z)$ around $z=1$
can be related to the LT critical limit of 
combinations of connected Green's functions evaluated at zero momentum 
\begin{equation}
\chi_{j} = \sum_{x_2,...,x_{j}}\langle \phi(0) \phi(x_2)
...\phi(x_{j-1}) \phi(x_{j})\rangle_c.
\label{chidef}
\end{equation}
In the LT critical limit
\begin{eqnarray}
&&- {\chi_3 M\over \chi^2}\longrightarrow  v_3,
\label{v3nc}\\
&&- {\chi_4 M^2\over \chi^3} + 3 {\chi_3^2 M^2\over \chi^4}
\longrightarrow  v_4,
\label{v4nc} \\
&&- {\chi_5 M^3\over \chi^4} + 
10 {\chi_4\chi_3 M^3\over \chi^5} - 
15 {\chi_3^3 M^3\over \chi^6}
\longrightarrow  v_5,
\label{v5nc} \\
&&- {\chi_6 M^4\over \chi^5} + 
15 {\chi_5\chi_3 M^4\over \chi^6} +
10 {\chi_4^2 M^4\over \chi^6} - 
105 {\chi_4\chi_3^2 M^4\over \chi^7} +
105 {\chi_3^4 M^4\over \chi^8}
\longrightarrow  v_6,
\label{v6nc} 
\end{eqnarray}
etc...,
where $M\equiv \chi_1$ and $\chi\equiv \chi_2$.

\section{The two-dimensional Ising model}
\label{sec3}

In the critical limit
the two-point function of the two-dimensional Ising model, 
both in the symmetric and broken phase, satisfies a 
Painlev\`e differential equation~\cite{Wuising}. Therefore,
$w$ can be calculated by an appropriate numerical integration, obtaining
\begin{equation}
w = 0.72906....
\label{wd2}
\end{equation}
The coefficients $v_j$ of the expansion of $B(z)$ are not known. Good
estimates of the first few $v_{j}$ can be obtained from the analysis of their
LT expansion.  The basic reason is that 
the leading correction to scaling is analytic, since
the subleading exponent $\Delta$ is expected to be larger 
than one (see e.g. Ref.~\cite{B-F} and references therein). 
In particular the available exact calculations~\cite{Wuising}
for the square-lattice Ising model near criticality 
have revealed only analytic corrections to the leading power law.
Therefore the traditional methods of series analysis should work well.

In order to estimate the ratios $v_j$ we used the results of 
Ref.~\cite{B-E}. They report the expansion of 
the free energy in the presence of an external magnetic field
to $O(u^{23})$ where $u\equiv e^{-4/T}$, for the square-lattice Ising model.
Using Eqs.~(\ref{v3nc},\ref{v6nc}) one can easily 
obtain the corresponding expansion for $v_j$.
In our analysis we used  several types of
approximants, 
Pad\'e, Dlog-Pad\'e and first-order integral 
approximants\footnote{
Given a $n$th order series,
we considered the following quasi-diagonal approximants:
$[l/m]$ Pad\'e and Dlog-Pad\'e approximants with
$l+m \geq n-2$ and
$l,m \geq  {n\over 2}-2$;
$[m/l/k]$ 
first-order integral approximants  with $m+l+k+2=n$ and
$\lfloor (n-2)/3 \rfloor -1\leq m,l,k \leq \lceil (n-2)/3\rceil +1$.
\label{ff}} 
(for a review on the resummation techniques
see for example Ref.~\cite{Guttrev}),
constructed from the series of $s_j\equiv u^{-2(j-2)}v_j^{-1}$
that have the form $\sum_{i=1}^{21} c_i u^i$.
Estimates of $v_j$ are then obtained by evaluating them
at $u_c=(\sqrt{2}+1)^{-2}$.  
In App.~\ref{appa} we report the $O(u^{23})$ series of the  
zero-momentum connected correlation functions $\chi_j$ (with $j\leq 6$) 
we used in our analysis.

We obtained~\footnote{
As estimate from each class of approximants, i.e. 
Pad\'e, Dlog-Pad\'e and first-order integral approximants,
we considered the average of the 
values at $u=u_c$  of the non-defective approximants 
using all the available terms of the series. As estimate of the
error we considered  
the square root of the variance around the estimate of the results 
from all the non-defective approximants listed in footnote \ref{ff}.
The results from Pad\'e, Dlog-Pad\'e and first-order integral approximants
were then combined leading to the estimates shown in the text.
\label{fff}}
\begin{eqnarray}
v_3 &=& 33.011(6), \label{d2v3} \\ 
v_4 &=& 48.6(1.2),  \label{d2v4} \\
v_5 &=& 7.69(2)\times 10^4,  \label{d2v5} \\
v_6 &\approx & -2.17\times 10^7 ,  \label{d2v6} \\
v_7 &\approx & 9.9\times 10^9 ,  \label{d2v7} \\
v_8 &\approx & -6.2\times 10^{12} .  \label{d2v8}
\end{eqnarray}
Notice how the absolute values of $v_j$ increase rapidly with $j$.
Using the value of $w$ reported in Eq.~(\ref{wd2}) we obtain:
\begin{eqnarray}
&&g_3 =wv_3=24.067(4),\label{g32d}\\
&&g_4 =w^2v_4=25.8(6).\label{g42d}
\end{eqnarray}
In Ref.~\cite{Z-L-F} the estimates 
$v_3=33.06(10)$ 
($v_3$ is denoted there by $R_3$) and $g_4 \approx 25$
(with a large apparent  uncertainty)
were obtained from the analysis of the LT series published
in Ref.~\cite{Sykesetal} for the square, 
triangular and honeycomb lattices.  
These results are fully consistent with our analysis.
The higher precision we achieved is essentially
due to the longer series we considered,
indeed for the square
lattice Ref.~\cite{Sykesetal} reports series to $O(u^{11})$.
We also mention the estimate $g_3=23.9(2)$ reported in 
Ref.~\cite{K-P} that has been obtained
by a Monte Carlo simulation for $\xi\lesssim 8$.

\section{The three-dimensional Ising model}
\label{sec4}

For the three-dimensional Ising model accurate estimates
of $w$ can be found in the literature.
They are obtained using various approaches,
from the analysis of low-temperature expansions~\cite{Z-L-F,ONgr2}, 
from Monte Carlo simulations~\cite{C-H},
and field-theoretic calculations~\cite{Munster,Munster2,B-T-W}.
The low-temperature expansion of $w^2$ can be calculated
to $O(u^{21})$ on the cubic lattice using the series published
in Refs.~\cite{Ar-Ta,Vohw}. In the analysis of this series we
used the Roskies transform~\cite{Roskies} in order
to reduce the systematic effects due to confluent singularities.
Indeed, since the leading correction-to-scaling exponent
$\Delta\simeq 1/2$,  the systematic
error due to the leading non-analytic correction may be relatively large.
The analysis using Pad\'e, Dlog-Pad\'e and first-order integral approximants
(see footnotes \ref{ff} and \ref{fff}) to the Roskies transformed series 
of $w^2$ leads to the estimate $w^2=4.75(4)$. This is in good agreement with
the estimate of $w^2$ that can be obtained from 
the results of Ref.~\cite{Z-L-F}, $w^2=4.71(5)$ \cite{Fisher-pr}. 
The result of the Monte Carlo simulation
reported in Ref.~\cite{C-H}, $w^2=4.77(3)$, is consistent with
these estimates. Therefore, we believe that
\begin{equation}
w=2.18(1)
\label{wd3}
\end{equation}
should be a reliable estimate.  In Refs.~\cite{Munster,Munster2} 
a perturbative approach in fixed dimension $d=3$ is exploited
to study the LT phase of the Ising model. The expansion parameter is 
$u\equiv 3w^2$. Its critical value 
is determined computing the zero of the corresponding
$\beta$-function that is known to three loops.
The estimate of $w^2$ one obtains from this expansion 
is consistent with Eq.~(\ref{wd3}), but less precise
due to the relatively small number of known term in the $\beta$-function.

In order to calculate the first few $v_j$ in three dimensions,
we consider their $\epsilon$-expansion.
The $\epsilon$-expansion of $v_j$ can be derived from the
$\epsilon$-expansion of the equation of state that is known to $O(\epsilon^3)$ 
for the Ising model~\cite{B-W-W-e2,W-Z,N-A}.
Using the procedure described in Sec.~\ref{sec2},
cf. Eqs.~(\ref{daz}) and (\ref{norm}),
one obtains
\begin{eqnarray}
v_3 &=& 3 + {3\over 2}\epsilon +
{17\over 18}\epsilon^2 + \left( {989\over 1944}
-{\lambda\over 12} -{\zeta(3)\over 6}\right)\epsilon^3 + 
O(\epsilon^4),\\
v_4 &=& 3 + {9\over 2}\epsilon +
{43\over 12}\epsilon^2 + \left( {1601\over 648}
+{\lambda\over 4} -{\zeta(3)\over 2}\right)\epsilon^3 + 
O(\epsilon^4),\\
v_5 &=& {9\over 2}\epsilon +
{31\over 3}\epsilon^2 + \left( {3919\over 324}
-{11\lambda\over 4} +{5\zeta(3)\over 2}\right)\epsilon^3 + 
O(\epsilon^4),\\
v_6 &=& -18 \epsilon -
{311\over 6}\epsilon^2 + \left( -{12167\over 162}
+{65\lambda\over 2} - 34\zeta(3)\right)\epsilon^3 + 
O(\epsilon^4),
\end{eqnarray}
etc..., where 
$\lambda=\psi'(1/3)/3 - 2\pi^2/ 9 \approx 
1.17195$, and $\zeta(3)\approx 1.20206$.

Since the $\epsilon$-expansion is asymptotic, 
it requires a resummation to get estimates at $d=3$,
i.e. $\epsilon=1$. Assuming, as usual, its Borel summability,
the analysis of the series is performed
using the method proposed in Ref.~\cite{LeG-ZJ}, that
is based on the knowledge of the large-order behaviour of the series.
Given a quantity $R$ with series 
\begin{equation}
R(\epsilon)=\sum_{k=0} R_k \epsilon^k,
\end{equation}
we have generated new series $R_p(\alpha,b;\epsilon)$ according to
\begin{equation}
R_p(\alpha,b;\epsilon) = \sum_{k=0}^p 
    B_k(\alpha,b) 
  \int^\infty_0 dt\ t^b\
  e^{-t} {u(\epsilon t)^k \over \left[1 - u(\epsilon t) \right]^\alpha}  ,
\label{RBorel}
\end{equation}
where 
\begin{equation}
   u(x) = { \sqrt{1 + a x} - 1\over \sqrt{1 + a x} + 1}.
\end{equation}
Here $a=1/3$ is the singularity of the Borel transform.
The coefficients $B_k(\alpha,b)$ are determined requiring
that the expansion in $\epsilon$ of $R_p(\alpha,b;\epsilon)$
coincides with the original series. For each $\alpha$, $b$ and $p$
an estimate of $R$ is simply given by $R_p(\alpha,b;\epsilon=1)$.
We follow Refs.~\cite{ONgr2,effpot} in order
to derive the estimates and their uncertainty.
We determine an integer value of $b$, $b_{\rm opt}$, such that
\begin{equation} 
R_3(\alpha,b_{\rm opt};\epsilon=1)\approx R_2(\alpha,b_{\rm opt};\epsilon=1)
\end{equation}
for $\alpha < 1$.
$b_{\rm opt}$ is the value of $b$ such that the estimate
from the series to order $O(\epsilon^3)$ is essentially identical 
to the estimate 
from the series to order $O(\epsilon^2)$. In a somewhat arbitrary way 
we  consider as our final estimate 
the average of $R_p(\alpha,b;\epsilon=1)$ with 
$-1 < \alpha \le 1$ and $-2 + b_{\rm opt} \le b \le 2 + b_{\rm opt}$.
The error we report is the variance of the values of
$R_3(\alpha,b;\epsilon=1)$
with $-1 < \alpha \le 1$ and 
$\lfloor b_{\rm opt}/3 - 1\rfloor \le b
     \le \lceil  4 b_{\rm opt}/3 + 1\rceil$.
A discussion of the reliability of these error estimates 
can be found in Refs.~\cite{ONgr2,effpot}.

When the coefficients of the $\epsilon$-expansion have all the same
sign, as it is the case for $v_j$, it is convenient to consider
and analyze the series of the inverse. 
The analysis of the $O(\epsilon^3)$ series
of $v_j^{-1}$ gives
\begin{eqnarray}
v_3 &=& 6.05(27), \label{v3st} \\
v_4 &=& 17.2(6.6). \label{v4st} 
\end{eqnarray}
The analysis of the $\epsilon$-series of the coefficients
$v_j$ with $j> 4$ is extremely unstable and does not provide 
reliable estimates.

These results can be improved  performing a constrained analysis
that exploits the accurate two-dimensional estimates
of $v_3$ and $v_4$, cf. Eqs.~(\ref{d2v3}) and (\ref{d2v4}).
The idea is the following. Assume that 
exact or approximate values of the quantity at hand 
are known for some dimensions
$d_i<3$, with $d_i$ belonging to the expected analytic domain in $d$.
Then one may use the polynomial interpolation between the
values $d=4$ and $d=d_i$ as zeroth order approximation 
in three dimensions. If the interpolation is a 
good approximation one should find that the series which gives the 
deviations has smaller coefficients than the original one. 
Consequently also the errors in the resummation are reduced.
The idea to constrain the $\epsilon$-series analysis was
employed in Ref.~\cite{eexp} to improve 
the estimates of the critical exponents of the Ising 
and self-avoiding walk models; in Ref.~\cite{greenfunc}
it was used in the study of the two-point function;
in Refs.~\cite{ONgr2,effpot} it was 
successfully applied to the study of the 
small-renormalized-field expansion of the effective
potential in the symmetric phase.
For quantities defined in the LT of the Ising model,
the analytic domain in $d$ should contain the value $d=2$. 
Thus we can use the rather accurate two-dimensional 
estimates of $v_3$ and $v_4$ to constrain the
analysis of the $\epsilon$-series.
In the study of the effective potential in the symmetric phase
also the exact results in $d=1,0$ were used, because in that
case the analytic domain is expected to extend up to $d=0$~\cite{ONgr2,effpot}.

Assuming that the ratios $v_{j}$ are analytic and sufficiently 
smooth in the domain $4>d\geq 2$ (that is $0 < \epsilon \leq 2$),
one may perform a linear interpolation between $d=4$ and $d=2$, 
and then analyze the series of the difference.
For a generic quantity $R$ one defines 
\begin{equation}
\overline{R}(\epsilon) = \left[ 
    {R(\epsilon) - R_{\rm ex}(\epsilon=2) \over (\epsilon - 2)}\right]
\end{equation}
and a new quantity 
\begin{equation}
   R_{\rm imp}(\epsilon) = R_{\rm ex}(\epsilon=2) + 
         (\epsilon - 2) \overline{R}(\epsilon),
\end{equation}
where $R_{\rm ex}(\epsilon=2)$ is the exact value of $R$ for $\epsilon=2$.
New estimates of $R$ for $\epsilon = 1$ can be obtained 
applying the resummation procedure we described above to 
$\overline{R}(\epsilon)$  and then computing 
$R_{\rm imp}(1)$. 

The constrained analysis of the $\epsilon$-series
of $v_j^{-1}$ leads to the following results:
\begin{eqnarray}
v_3 &=& 5.99(5+0), \label{v3im} \\
v_4 &=& 15.8(1.4+0.1), \label{v4im} 
\end{eqnarray}
where the second error is obtained varying the two-dimensional 
estimate within one error bar. The new results are in good 
agreement with the estimates  (\ref{v3st}) and (\ref{v4st}), but 
have a smaller uncertainty. Using the estimate (\ref{wd3}) of
$w$ we can compute the three- and four-point
zero-momentum renormalized coupling constants:
\begin{eqnarray}
g_3  &=& 13.06(12), \label{g3} \\
g_4  &=& 75(7). \label{g4} 
\end{eqnarray}
The errors of $g_3$ and $g_4$ are calculated considering
the errors of $w$, $v_3$, and $v_4$ as independent.

Let us compare our results with available estimates obtained using 
other methods.  Table~\ref{gdata} presents a summary of all 
the available (as far as we know) estimates of $v_3$, $g_3$ and $g_4$.
Field-theoretic estimates of $v_3$ are obtained in Refs.~\cite{G-Z,G-Z-2}  
using the parametric representation of the equation  of state
(in Table~\ref{gdata} we refer to this approach by PR).
In this approach the $\epsilon$-expansion and the $d=3$ $g$-expansion (i.e.
expansion in powers of the HT zero-momentum four-point
renormalized coupling) are used to estimate the first few 
coefficients of the expansion in powers of $\theta$ of the function
$h(\theta)$ characterizing
the parametric representation of the equation of state. From $h(\theta)$ 
one can derive many universal ratios of quantities
defined at zero momentum such as $v_j$.
Ref.~\cite{Z-L-F} presents an analysis of the low-temperature expansion
on the cubic, b.c.c. and f.c.c. lattices, obtaining
$v_3=6.47(20)$, and $g_4 \approx 85$ with a large
apparent uncertainty. These estimates are in good agreement 
with our results (\ref{v3im}) and (\ref{g4}).
We also mention Ref.~\cite{B-T-W} where the effective
potential in the broken phase is determined from an approximate solution
of the exact renormalization-group equations.
The resulting estimates are $w^2=5.55$ and $g_3=15.24$. 
It is very hard to obtain good estimates by Monte Carlo simulations.
Ref.~\cite{K-P} reports some results obtained with a Monte Carlo 
simulation with $\xi\lesssim 8$. They obtain $g_3 =16.0(5)$,
and $g_4-3g_3^2\approx 750$ (the precision
of the data does not allow us to extract an estimate of $g_4$).
Ref.~\cite{Tsypinbr} reports estimates of $g_3$ and $g_4$ that are obtained 
from the probability distribution of the average magnetization,
whose numerical data  are analyzed by assuming a  
sixth-order polynomial approximation of the effective potential.
In  Table \ref{gdata} we report the data of Ref.~\cite{Tsypinbr} 
corresponding to  the largest correlation length
$\xi\simeq 7$ and lattice  $74^3$ (taken
from the 5th column of Table 2 as suggested by the author).
While for $g_3$ there is a substantial agreement, 
the estimate of $g_4$ turns out to be quite larger 
than ours.

\section{The effective potential in the large-$N$ limit}
\label{sec5}

The physics of the broken phase of O($N$) models
with $N>1$ is very different from that of the Ising model, because
of the presence of Goldstone modes.
In this case the effective potential 
${\cal F}(\varphi)$ is not analytic for $\varphi\to\varphi_0$.
General renormalization-group arguments predict
\begin{equation}
{\cal F}(\varphi) - {\cal F}(\varphi_0)
\approx c\left( \varphi^2-\varphi_0^2\right)^{\rho} ,
\label{freeengbg}
\end{equation}
where $c$ is a constant and $\rho$ a new exponent that is 
conjectured to be given exactly by $\rho = d/(d-2)$.  
The conjecture is based on the following argument. 
The exponent $\rho$ is related to the behaviour of the 
longitudinal susceptibility $\chi_L$ along the coexistence curve. From 
Eq. (\ref{freeengbg}) it is easy to derive 
\begin{equation}
\chi_L  = {\partial M\over \partial H} \sim H^{(2-\rho)/(\rho-1)},
\end{equation}
for $H\to 0$. On the other hand the singularity of $\chi_L$ for $H\to0$
is governed by the zero-temperature infrared-stable fixed
point~\cite{B-W,B-Z,Lawrie}.  This leads to the exact prediction 
\begin{equation}
\chi_L \sim H^{d/2-2},
\label{chil}
\end{equation}
and therefore $\rho = d/(d-2)$. The asymptotic behaviour (\ref{freeengbg}) 
has been checked in the large-$N$ limit to leading and 
subleading order \cite{B-W}. At leading order
Eq. (\ref{freeengbg}) holds for all $\varphi$ and not only in the limit 
$\varphi\to\varphi_0$.

The nature of the corrections to the behaviour 
(\ref{freeengbg}) is less clear. Setting 
$x = t M^{-1/\beta}$ and $y=H M^{-\delta}$, the conjecture is that 
$1+x$ has the form of a double expansion in powers of $y$ and $y^{(d-2)/2}$
near the coexistence curve~\cite{W-Z-2,S-H,Lawrie}, i.e.
for $y\to0$ 
\begin{equation}
1+x = c_1 y + c_2 y^{1-\epsilon/2} + d_1 y^2 + 
       d_2 y^{2-\epsilon/2} + d_3 y^{2-\epsilon} + \ldots
\label{xp1intermsofy}
\end{equation}
where $\epsilon = 4 - d$. Here we assume, as usual, 
that $x=-1$ corresponds to the coexistence curve. With this normalization 
$z = (-x)^{-\beta}$. Thus, for $x\to -1$, $1+x \approx (z-1)/\beta$ with
corrections that are analytic in $(z-1)$. Therefore, also $z-1$ should have 
an expansion of the same form.

Notice that this expansion predicts that, in three dimensions,
$z-1$ has an expansion that is analytic in powers of $y^{1/2}$. 
As a consequence, the effective potential 
would be analytic in $z-1$, with $B(z) \sim (z-1)^3$ for 
$z\to 1$.

These ideas can be verified in the large-$N$ limit using 
the equation of state to order $O(1/N)$ (next-to-leading in $1/N$) 
reported in Ref.~\cite{B-W}. Setting $\omega = 1 + x $, we find 
that, in generic dimension $d$ with $2<d<4$, the function $f(x)$ defined in Eq. 
(\ref{eqstate}) has an expansion of the form
\begin{equation}
f(x) =  \omega^{2/(d-2)}\left[ 
1 + {1\over N}\left( \sum_{i=0}^\infty c_i \omega^i
    + \sum_{n=-1}^\infty \sum_{m=1}^\infty c_{mn} \omega^{2m/(d-2)+n} \right) + 
       O\left( {1\over N^2} \right)\right],
\label{bzlnasy}
\end{equation}
which is consistent with the expansion (\ref{xp1intermsofy}).
Generically, the coefficients $c_{mn}$ are singular for those values of $d$
such that $2m/(d-2)$ is an integer, i.e. for
\begin{equation}
2 < d = 2 + {2 m\over n} < 4, \qquad {\rm for}\quad 
0< m < n,\quad  m,n\in\ \N.
\label{speciald}
\end{equation}
However, in these cases, also
the coefficients $c_i$ are singular. A careful analysis shows 
that for these special values of $d$ logarithmic terms should appear. 
We have verified analytically their presence for 
$d = 4 - 2/n$, that corresponds to
$m=n-1$ in Eq.~(\ref{speciald}). Explicitly in three dimensions ($n=2$) we find
\begin{equation}
f(x) = \omega^2\left[1 + {1\over N}\left(
  f_1(\omega) + \log \omega f_2(\omega) \right)
    + O(N^{-2})\right] \; .
\label{fx3d}
\end{equation}
The functions $f_1(\omega)$ and $f_2(\omega)$ have a regular
expansion in powers of $\omega$. In particular
\begin{equation}
f_2(\omega) = - {23\over6} \omega^2 - 
    {1\over 16} (128 + \pi^2) \omega^3 + 
    {1\over 160} (2000 + 17 \pi^2) \omega^4 + O(\omega^5).
\end{equation}
Notice the absence of a linear term in $f_2(\omega)$.
This is due to the fact that $c_{1,-1}$ and $c_1$ in
Eq. (\ref{bzlnasy}) are not singular for $d\to 3$.

The logarithmic terms we find are not predicted by Eq.
(\ref{xp1intermsofy}). In particular Eq.~(\ref{fx3d}) 
is incompatible with Eq. (\ref{xp1intermsofy}),
that, as we already remarked, predicts an analytic expansion in powers
of $\omega$. 
The presence of logarithms in the expansion for values of $d$ arbitrarily 
near 4 casts some doubts on the asymptotic behavior (\ref{xp1intermsofy}) that 
has been derived essentially from an $\epsilon$-expansion analysis.

What do we learn for the behaviour of the critical effective potential 
near the coexistence curve for finite values of $N$ ? 
A possible interpretation of the 
results is that the expansion (\ref{fx3d}) holds for all values of $N$ and 
thus Eq. (\ref{xp1intermsofy}) is correct apart from logarithms that 
are present for some special values of $d$. 
The reason of their appearance is however unclear. 
Neverthless, it does not necessarily contradict
the conjecture that the behavior near the coexistence curve
is controlled by the zero-temperature 
infrared-stable Gaussian fixed point. 
In this case logarithms would not be unexpected,
as they usually appear in the reduced temperature
asymptotic expansion around Gaussian fixed points
(see e.g. Ref.~\cite{B-B}).
Anyway,
since they appear also for dimensions $d > 3$, they cannot be related
to the presence of marginal operators, say $(\varphi^2)^3$ in three dimensions. 

There is another possible interpretation of the result (\ref{fx3d})
that we mention for completeness.
In the large-$N$ expansion, logarithmic terms are usually the signal 
of the presence of $N$-dependent critical exponents. For instance consider 
a term of the form
\begin{equation}
A(N) \omega^{\sigma(N)} + B(N) \omega^m,
\label{sigm}
\end{equation}
with $\sigma(N) = m + \sigma_1/N + O(N^{-2})$ and $m$ integer.
For large values of $N$, expanding $A(N) = A_0 + A_1/N +O(N^{-2})$ 
and analogously $B(N)$, we obtain
\begin{equation}
(A_0 + B_0) \omega^m + 
   {1\over N}\left[(A_1 + B_1)\omega^m + 
    \sigma_1 A_0 \omega^m \log \omega\right] + O(N^{-2}).
\end{equation}
This expansion would reproduce the behaviour (\ref{fx3d}), provided
$A_0 + B_0 = 0$. In generic dimension Eq. (\ref{sigm}) would 
be compatible with the expansion (\ref{bzlnasy}) if 
\begin{equation}
  \sigma = {2 k\over d-2} + m - 2 k + O(N^{-1}),
\end{equation}
for some integer $k \le m/2$. Notice that the expansion in 
powers of $1/N$ of the coefficients $A(N)$
and $B(N)$ would be discontinuous in $d$, since $A(N)$ and $B(N)$
would be of order $1/N$ in generic dimension and of order
$1$ whenever $\sigma(\infty)$ is an integer. A similar 
phenomenon has been found for the HT specific heat 
\cite{Abe-Hikami,Abe-Hikami-2}, and in the expansion of the 
Callan-Symanzik $\beta$-function near the critical point \cite{ONgr2}.
This interpretation of the results 
is quite natural in the framework of the large-$N$ expansion. 
However it is not clear which exponents should appear. In particular, 
if the singular behaviour is controlled by the zero-temperature 
infrared-stable Gaussian fixed point, we would not expect any $N$-dependent 
exponent. 
We think that these questions deserve further investigation.

\acknowledgments

We acknowledge many useful discussions with Paolo Rossi.

\appendix

\section{Two-dimensional LT Series}
\label{appa}

In this appendix we report the LT expansions of the 
connected correlation functions $\chi_j$ for the the square-lattice Ising model
with nearest-neighbor interactions. They are derived from the 
expansion of the free energy in the presence of an external magnetic 
field~\cite{B-E}.
This series reproduces the exact results for the 
free energy and the magnetization in the absence of a magnetic field
\cite{Onsager,Baxter}
\begin{eqnarray}
F &= & \int_0^\pi {d\theta\over 2\pi}
 \log\left\{ {(1+u)^2\over 2u} + {1\over 2u}
   \left[ (1 + u)^4 - 16 u (1-u)^2 \sin^2 \theta\right]^{1/2}\right\} , 
\label{FH02d}\\
M &=& \left[ 1 - {16 u^2\over (1-u)^4}\right]^{1/8},
\label{MH02d}
\end{eqnarray}
where $u = e^{-4/T}$. This check is particularly effective as all 
the terms of the expansion contribute in the limit $H\to 0$.

The expressions for the susceptibilities, cf. Eq. (\ref{chidef}),
up to terms of order $O(u^{24})$ are the following:
\begin{eqnarray}
\hskip -10pt
\chi &=& u^2(4 + 32 u + 240 u^{2} + 1664 u^{3} + 11164 u^{4} + 73184 u^{5} + 
    472064 u^{6} + 3008032 u^{7} 
\nonumber \\
\hskip -5pt && \hskip -12pt
   + 18985364 u^{8} + 118909888 u^{9} + 740066448 u^{10} + 
   4581660832 u^{11}
\nonumber \\
\hskip -5pt && \hskip -12pt
   + 28237063308 u^{12} + 173353630848 u^{13} + 
   1060674765568 u^{14} 
\nonumber \\
\hskip -5pt && \hskip -12pt
   + 6470624695296 u^{15} + 39370663086596 u^{16} + 
   238993166711328 u^{17} 
\nonumber \\
\hskip -5pt && \hskip -12pt
   + 1447734754083760 u^{18} + 8753312020985216 u^{19} + 
   52833859249062236 u^{20} 
\nonumber \\
\hskip -5pt && \hskip -12pt
   + 318401517346022240 u^{21})
\\
\hskip -10pt
\chi_3 &=& u^2(-8 - 128 u - 1648 u^{2} - 17216 u^{3} - 161608 u^{4} - 
   1407040 u^{5} - 11607872 u^{6} 
\nonumber \\
\hskip -5pt && \hskip -12pt
   - 91914944 u^{7} - 704714056 u^{8} - 5264048832 u^{9} - 
   38484337136 u^{10} 
\nonumber \\
\hskip -5pt && \hskip -12pt
   - 276313973760 u^{11}- 1953615567368 u^{12} - 
   13630566361728 u^{13} 
\nonumber \\
\hskip -5pt && \hskip -12pt
   - 94009039294720 u^{14} - 641820330787712 u^{15} - 
   4342587216599624 u^{16} 
\nonumber \\
\hskip -5pt && \hskip -12pt
   - 29147168513086720 u^{17} - 194228811717309808 u^{18} - 
   1285893394538102720 u^{19} 
\nonumber \\
\hskip -5pt && \hskip -12pt
   - 8463157853682576776 u^{20} - 
   55401729118133053376 u^{21})
\\
\hskip -10pt
\chi_4 &=& u^2(16 + 512 u + 11232 u^{2} + 174464 u^{3} + 2268016 u^{4} + 
   26007680 u^{5} 
\nonumber \\
\hskip -5pt && \hskip -12pt
   + 272557376 u^{6} + 2666887552 u^{7} + 24721765328 u^{8} + 
   219362198656 u^{9} 
\nonumber \\
\hskip -5pt && \hskip -12pt
   + 1877373570336 u^{10} + 15586471370752 u^{11}+ 126094785804720 u^{12} 
\nonumber \\
\hskip -5pt && \hskip -12pt
   + 997562812048896 u^{13} + 7739640438305152 u^{14} + 
   59027832457279488 u^{15} 
\nonumber \\
\hskip -5pt && \hskip -12pt
   + 443396861865190928 u^{16} + 3285761490274027008 u^{17} + 
   24053962791473093728 u^{18} 
\nonumber \\
\hskip -5pt && \hskip -12pt
   + 174163313822385408896 u^{19} + 
   1248496572536156377712 u^{20} 
\nonumber \\
\hskip -5pt && \hskip -12pt
   + 8868728488366239392384 u^{21})
\\ 
\hskip -10pt
\chi_5 &=& u^2(-32 - 2048 u - 76864 u^{2} - 1763072 u^{3} - 31561504 u^{4} - 
   474188032 u^{5} 
\nonumber \\
\hskip -5pt && \hskip -12pt
   - 6284096768 u^{6} - 75680753408 u^{7} - 845304372256 u^{8} - 
   8883063406848 u^{9} 
\nonumber \\
\hskip -5pt && \hskip -12pt
   - 88761722644544 u^{10} - 850124864318976 u^{11}- 
   7853055915373856 u^{12} 
\nonumber \\
\hskip -5pt && \hskip -12pt
   - 70313539072037376 u^{13} - 612653077144892416 u^{14} - 
   5211746985723756032 u^{15} 
\nonumber \\
\hskip -5pt && \hskip -12pt
   - 43403126238486164768 u^{16} - 
   354661581988366266880 u^{17} 
\nonumber \\
\hskip -5pt && \hskip -12pt
   - 2849039732654299216960 u^{18} - 
   22536604900114543443200 u^{19} 
\nonumber \\
\hskip -5pt && \hskip -12pt
   - 175792207497016658107424 u^{20} - 
   1353835890235723182430976 u^{21})
\\
\hskip -10pt
\chi_6 &=&
  u^2(64 + 8192\,u + 531840\,{u^2} + 17944064\,{u^3} + 440964544\,{u^4} + 
   8650801664\,{u^5} 
\nonumber \\
\hskip -5pt && \hskip -12pt
   + 144516779264\,{u^6} + 2136034381312\,{u^7} + 
   28672334567744\,{u^8} 
\nonumber \\
\hskip -5pt && \hskip -12pt
   + 356014396523008\,{u^9} + 
   4144706261854848\,{u^{10}} + 45707724350236672\,{u^{11}} 
\nonumber \\
\hskip -5pt && \hskip -12pt
   + 481293469566545088\,{u^{12}} + 4869609919772817408\,{u^{13}} + 
   47583052907024977408\,{u^{14}} 
\nonumber \\
\hskip -5pt && \hskip -12pt
   + 450914521123508084736\,{u^{15}} + 
   4158352425040048310336\,{u^{16}} 
\nonumber \\
\hskip -5pt && \hskip -12pt
   + 37427697740234745151488\,{u^{17}} + 
   329592281476942345924480\,{u^{18}} 
\nonumber \\
\hskip -5pt && \hskip -12pt
   + 2845676740202773130026496\,{u^{19}} +
   24132653400815900244380096\,{u^{20}} 
\nonumber \\
\hskip -5pt && \hskip -12pt
   + 201334455541709269472268800\,{u^{21}})
\end{eqnarray}
We should note that the expansion of $\chi$ does not reproduce 
the expression reported in Ref. \cite{B-E}. The last two terms 
are slightly different:
\begin{equation}
 \chi({\rm our}) - \chi({\rm Ref. \protect\cite{B-E} }) = 
   52 u^{22} + 872 u^{23}
\end{equation}
The discrepancy is probably due to a misprint, since
their general series (in $\mu=e^{-2 H}$ and $u$) correctly 
reproduces Eqs. (\ref{FH02d},\ref{MH02d}).


\begin{table}
\caption{We report the available results for $v_3$, $g_3$ and $g_4$.
The estimates of $g_3$ marked by an asterisk
have been derived by us using the value of $v_3$ of the corresponding
line and the estimate (\ref{wd3}) of $w$.
\label{gdata}}
\begin{tabular}{cr@{}lr@{}lr@{}l}
\multicolumn{1}{c}{}&
\multicolumn{2}{c}{$v_3$}&
\multicolumn{2}{c}{$g_3$}&
\multicolumn{2}{c}{$g_4$}\\
\tableline \hline
$\epsilon$-exp. [this paper] & 5&.99(5)  & 13&.06(12) & 75&(7) \\
$\epsilon$-exp. PR~\cite{G-Z-2} & 6&.07(17) & $*$13&.2(4) &  & \\
$d=3$ -exp. PR~\cite{G-Z-2} & 6&.08(6) & $*$13&.25(14)& & \\
ERG~\cite{B-T-W} & 6&.47 & 15&.24 & & \\
LT~\cite{Z-L-F}& 6&.47(20) & 13&.9(4) & $\approx$ 85& \\
MC~\cite{Tsypinbr} & & & 13&.6(5) & 108&(7) \\
\end{tabular}
\end{table}

\end{document}